# Single-molecular and Ensemble-level Oscillations of Cyanobacterial Circadian Clock

Sumita Das[1], Tomoki P. Terada[1], Masaki Sasai[1] *

[1]Department of Computational Science and Engineering and
Department of Applied Physics, Nagoya University, Nagoya 464-8603, Japan

Running Title:
Oscillations of Cyanobacterial Circadian Clock

Correspondence:
*email: sasai@nuap.nagoya-u.ac.jp
 telephone: +81-52-789-4763

## Abstract (less than 250 words)

When three cyanobacterial proteins, KaiA, KaiB, and KaiC, are incubated with ATP in vitro, the phosphorylation level of KaiC hexamers shows stable oscillation with approximately 24 h period. In order to understand this KaiABC clockwork, we need to analyze both the macroscopic synchronization of a large number of KaiC hexamers and the microscopic reactions and structural changes in individual KaiC molecules. In the present paper, we explain two coarse-grained theoretical models, the many-molecule (MM) model and the single-molecule (SM) model, to bridge the gap between macroscopic and microscopic understandings. In the simulation results with these models, ATP hydrolysis drives oscillation of individual KaiC hexamers and ATP hydrolysis is necessary for synchronizing oscillations of a large number of KaiC hexamers. Sensitive temperature dependence of the lifetime of the ADP bound state in the CI domain of KaiC hexamers makes the oscillation period temperature insensitive. ATPase activity is correlated to the frequency of phosphorylation oscillation in the single molecule of KaiC hexamer, which should be the origin of the observed ensemble-level correlation between the ATPase activity and the frequency of phosphorylation oscillation. Thus, the simulation results with the MM and SM models suggest that ATP hydrolysis randomly occurring in each CI domain of individual KaiC hexamers is a key process for oscillatory behaviors of the ensemble of many KaiC hexamers.

## Key Words (up to 5 items)

KaiABC, ATP hydrolysis, Synchronization, Temperature compensation, Coarse-grained models

## Significance Statement (less than 100 words)

Cyanobacterial proteins, KaiA, KaiB, and KaiC, can reconstitute a circadian clock when they are incubated with ATP in vitro. In order to understand this prototypical oscillator, we need to analyze both synchronization of a macroscopically large number of oscillating molecules and microscopic reactions in individual molecules. We introduced two theoretical models to unify macroscopic and microscopic viewpoints. Simulation results suggest that ATP hydrolysis is necessary for synchronization and temperature compensation and that ATPase activity is correlated to the oscillation frequency in individual molecules. Thus, ATP hydrolysis randomly occurring in individual molecules should determine important features of the ensemble-level oscillation.

# Introduction

A cyanobacterial protein, KaiC, shows a stable oscillation in its phosphorylation level with approximately 24 h period when three proteins, KaiA, KaiB, and KaiC, are incubated with ATP *in vitro* [1]. Much attention has been focused on this prototypical oscillator and close investigations have been made on interactions among Kai proteins [2, 3]. As shown in Fig. 1, KaiC monomer consists of tandemly repeated N-terminal (CI) and C-terminal (CII) domains [4] and assembles into a hexamer [5, 6] forming the CI-CII double rings [7]. We denote this KaiC hexamer as $C_6$. KaiA forms a dimer [8] and binds to the CII ring of $C_6$ to form a complex, $C_6A_2$ [9]. KaiB binds to the CI of each subunit of KaiC hexamer, and a KaiA dimer can further bind to each KaiB to form $C_6B_6A_{2j}$ with $0 \leq j \leq 6$ [10]. KaiC shows auto-phosphorylation (P) and auto-dephosphorylation (dP) at two specific sites, Ser431 and Thr432, of each CII domain, and the P reactions are promoted in the $C_6A_2$ complex [11-13]. In $C_6B_6A_{2j}$, binding of KaiA to the CII is suppressed, which promotes the dP reactions of KaiC [14, 15]. Thus, the temporal appearance of different kinds of complexes, accompanied by the switching between P-phase and dP-phase, is the important feature of the oscillation in the KaiABC system.

In order to elucidate the mechanism of this KaiABC clockwork, we need to understand both the microscopic atomic-scale dynamics of reactions in individual molecules [16] and the macroscopic ensemble-level synchronization among many KaiC molecules; because the ensemble-level oscillation vanishes when individual molecules oscillate independently of each other, synchronization of a macroscopically large number of molecules is necessary for maintaining the coherent oscillation as observed in test tube. Therefore, it is important to understand mechanisms at both microscopic and macroscopic levels.

A key observation for solving the problem of the relationship between mechanisms at two levels is the slow ATPase reactions in KaiC. The CI domain of each KaiC subunit hydrolyzes approximately 10 ATP molecules and the CII domain of each subunit hydrolyzes several ATP molecules per one day [16, 17]. In particular, by comparing KaiC molecules of various mutants, it was shown that the ATPase activity is correlated to the frequency of the ensemble-level P/dP rhythm [16, 17], where the ATPase activity was measured by the frequency of ADP release in the condition in the absence of KaiA and KaiB, which is the non-oscillatory condition, and the frequency of the ensemble-level P/dP rhythm was measured in the oscillatory condition with appropriate concentrations of KaiA and KaiB. Furthermore, the ATPase activity of the truncated CI domain shows correlation to the frequency of the ensemble-level P/dP rhythm [16]. These observations suggested the intrinsic relationship between ATP hydrolysis occurring in the CI of individual molecules and the P/dP rhythm of the ensemble of many molecules [16-18].

Many theoretical models were developed by describing concentrations of various molecular species with continuous variables in kinetic differential equations to model the ensemble-level macroscopic dynamics of the oscillating system [11, 19-24]. With these macroscopic models, various nonlinear mechanisms, including KaiA sequestration at various phases [11, 19-23] and monomer shuffling [24], were proposed to explain how the oscillation is maintained. Other models described the system with the Monte Carlo methods by explicitly considering stochastic P/dP reactions in individual molecules and stochastic binding reactions among many molecules [23-27]. These Monte Carlo-type models should be useful for analyzing the relation between microscopic and macroscopic mechanisms. However, the ATP hydrolysis reactions and the activated intramolecular dynamics with ATP consumption were not considered in these models. The role of ATP consumption was analyzed by recent models [28, 29], and the further comprehensive analyses are required for elucidating the relationship between ATP hydrolysis and the P/dP rhythm [29].

Here, in the present paper, we explain two "mesoscopic" models of oscillation for bridging the microscopic and macroscopic-level descriptions; the many-molecule (MM) model, and the single-molecule (SM) model. In both models, structure and reactions in individual KaiC molecules are represented by coarse-grained variables. For example, structure of the $k$th KaiC hexamer is represented by an order parameter, $X_k$ ($X$ in the SM model by dropping the suffix $k$); structural transition is described by the change between states with $X_k \approx 0$ and $X_k \approx 1$. In particular, the role of ATP hydrolysis is investigated with these models by considering the effects of ATPase reactions on the structural change; $X_k$ is perturbed by the ATP hydrolysis in these models, and binding/dissociation reactions, P/dP reactions, and ATPase reactions are affected by $X_k$. With the MM model in the present paper, intermolecular interactions are analyzed to examine the necessary conditions for the stable ensemble-level oscillation. This model was introduced in our previous publication [29], and is used in the present paper to analyze the mechanism of synchronization and temperature compensation. In the SM model, more emphasis is placed on the intramolecular dynamics to analyze the origin of oscillation of individual molecules. In the present paper, we discuss the linkage between microscopic and macroscopic oscillations through the combined analyses with the MM and SM models.

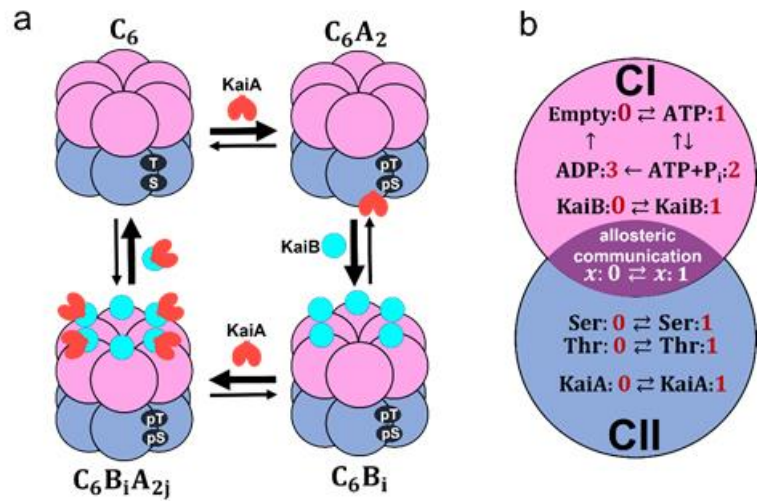

**Figure 1.** Coarse-grained models of the KaiABC system. **a**) Schematic of interactions among Kai proteins considered in the MM model. Two phosphorylation sites, Ser431 and Thr432, of each subunit in the CII domain (blue) are repeatedly phosphorylated (pS, pT) and dephosphorylated (S, T). KaiA dimer (red) binds to KaiC hexamer $C_6$ to form $C_6A_2$. The CI (pink) of each subunit binds

to KaiB (cyan) to form $C_6B_i$ with $1\leq i \leq 6$, which further binds to KaiA dimers to form $C_6B_iA_{2j}$ with $1\leq j \leq i$. **b)** Reactions and states in a KaiC subunit in the SM model. ATP hydrolysis reactions and KaiB binding/dissociation reactions occur in the CI domain (pink). Phosphorylation (P)/dephosphorylation (dP) reactions and KaiA binding/dissociation reactions take place in the CII domain (blue). CI and CII domains are coupled through allosteric structural change.

## Methods
**The many-molecule (MM) model.**
In order to analyze the oscillation of ensemble of molecules, we use the many-molecule (MM) model introduced in our previous publication [29]. In the MM model, we describe structure and reactions of individual molecules in a coarse-grained manner and calculate the coupled dynamics of $N = 1000$ hexamers of KaiC by explicitly following the dynamics of individual hexamers. As represented in Fig. 1a, we assume that a KaiA dimer binds to a KaiC hexamer to form $C_6A_2$. KaiB monomer binds to each subunit of KaiC to form $C_6B_i$ with $1 \leq i \leq 6$, and KaiA can further bind to $C_6B_i$ to form $C_6B_iA_{2j}$ with $1 \leq j \leq i$ [10]. These binding/unbinding reactions are represented as

$$2A + C_6 \rightleftharpoons C_6A_2,$$

$$A + C_6B_iA_{2j} \rightleftharpoons C_6B_{i+1}A_{2j}, \quad \text{for } 0 \leq i \leq 5, \ 0 \leq j \leq i,$$

$$2A + C_6B_iA_{2j} \rightleftharpoons C_6B_iA_{2j+2}, \quad \text{for } 1 \leq i \leq 6, \ 0 \leq j \leq i-1. \quad (1)$$

We should note that $C_6B_0A_0$ in Eq. 1 is identical to $C_6$. We define $P_{C_6A_2}(k,t)$ and $P_{C_6B_iA_{2j}}(k,t)$ as the probabilities of the $k$th KaiC hexamer $C_6$ to be bound in forms of $C_6A_2$ and $C_6B_iA_{2j}$, respectively, with $0 \leq i \leq 6$ and $0 \leq j \leq i$ at time $t$ and $k = 1, \ldots, N$. Thus, the kinetic equations for $P_{C_6A_2}$ and $P_{C_6B_iA_{2j}}$ are given as

$$\frac{d}{dt} P_{C_6A_2}(k,t) = h_A A^2 P_{C_6B_0A_0}(k,t) - f_A P_{C_6A_2}(k,t), \quad (2)$$

$$\frac{d}{dt} P_{C_6B_iA_{2j}}(k,t) = (7-i)h_B B P_{C_6B_{i-1}A_{2j}}(k,t) - (i-j)f_B P_{C_6B_iA_{2j}}(k,t)$$

$$- (6-i)h_B B P_{C_6B_iA_{2j}}(k,t) + (i+1-j)f_B P_{C_6B_{i+1}A_{2j}}(k,t)$$

$$+ (i-j+1)h_{BA} A^2 P_{C_6B_iA_{2j-2}}(k,t) - jf_{BA} P_{C_6B_iA_{2j}}(k,t)$$

$$- (i-j)h_{BA} A^2 P_{C_6B_iA_{2j}}(k,t) + (j+1)f_{BA} P_{C_6B_iA_{2j+2}}(k,t). \quad (3)$$

Here, $h_A, f_A, h_B, f_B, h_{BA}$ and $f_{BA}$ are rate constants, and $A$ and $B$ are concentrations of unbound free KaiA and KaiB molecules, respectively. When we numerically solve Eqs. 2 and 3, we impose the following constraints;

$$P_{C_6A_2}(k,t) + \sum_{i=0}^{6} \sum_{j=0}^{i} P_{C_6B_iA_{2j}}(k,t) = 1, \quad \text{for } k = 1, \ldots, N,$$

$$A_T = A + \frac{2}{V} \sum_{k=1}^{N} \left( P_{C_6A_2}(k,t) + \sum_{i=1}^{6} \sum_{j=1}^{i} j P_{C_6B_iA_{2j}}(k,t) \right),$$

$$B_T = B + \frac{1}{V} \sum_{k=1}^{N} \sum_{i=1}^{6} \sum_{j=1}^{i} i P_{C_6B_iA_{2j}}(k,t), \quad (4)$$

where $A_T$ and $B_T$ are total concentrations of KaiA and KaiB, respectively, and $V$ is the volume of the system. In this paper, we use units with $V = 1$.

Biochemical and structural observations [30-33] have suggested that the structure of KaiC hexamer at the phosphorylated state is different from the structure of KaiC hexamer at the unphosphorylated state. The structure at the unphosphorylated state is more stable than the structure at the phosphorylated state [33]. Assuming that a KaiC hexamer has a tight structure at the unphosphorylated state and a loose structure at the phosphorylated state, we can represent the allosteric transition between tight and loose structural states with a continuous variable $X$ as $X_k \approx 1$ when the structure of the $k$th KaiC hexamer is in the tight structural state, and $X_k \approx 0$ when it is in the loose structural state. In Eqs. 2 and 3, we assume that the binding rate constants, $h_A$ and $h_B$, and the dissociation rate constants, $f_A$ and $f_B$, depend on the KaiC structure, $X_k(t)$;

$$h_A(k,t) = h_A^1 X_k, \quad f_A(k,t) = f_A^0(1 - X_k),$$
$$h_B(k,t) = h_B^0(1 - X_k), \quad f_B(k,t) = f_B^1 X_k, \qquad (5)$$

where $h_A^1, h_B^0, f_A^0$, and $f_B^1$ are constants. With this assumption, the binding affinity of KaiA to KaiC increases when the $k$th KaiC is in the tight structural state. On the contrary, the loose structure of KaiC promotes KaiB binding to the KaiC hexamer. Because KaiA binds to KaiB in making $C_6B_iA_{2j}$, and does not bind directly to KaiC, we assume that the binding rate constant, $h_{BA}$, and the dissociation rate constant, $f_{BA}$, are independent of $X_k(t)$.

In the MM model, the phosphorylation level of twelve sites (Ser431 and Thr432 in six subunits) of a single KaiC hexamer are represented by a continuous variable $D$ as $D_k(t) \approx 1$ when the 12 sites of the $k$th KaiC hexamer are fully phosphorylated, and $D_k(t) \approx 0$ when the 12 sites are fully dephosphorylated. Temporal change in the phosphorylation level is represented as

$$\frac{dD_k}{dt} = k_p P_{C_6A_2}(k,t) - k_{dp}\left(1 - P_{C_6A_2}(k,t)\right) - \frac{\partial}{\partial D_k}g(D_k). \qquad (6)$$

Here, $k_p$ and $k_{dp}$ are rate constants of P and dP reactions, respectively. The term $g(D) = aD(D-1)(D-1/2)^2$ with $a > 0$ gives a constraint that confines the value of $D$ within a finite range. We also assume that the structural transition of KaiC between tight and loose states is much faster than other reactions in the system; therefore, the structure $X_k$ can be represented as a quasi-equilibrium average. In this paper, this average is determined under the mean-field generated by $D_k$, the binding probability of KaiA and KaiB, and the effect of ATPase reaction as

$$X_k(t) = \frac{1}{2}\tanh\left[\beta\left(c_0 - c_1 D_k(t) + c_2 p_k^A(t) - c_3 p_k^B(t) - q_k(t)\right)\right] + \frac{1}{2}, \qquad (7)$$

where $\beta = 1/k_B T$ and $c_0, c_1, c_2$, and $c_3$ are constants. $p_k^A(t) = P_{C_6A_2}(k,t)$ represents the level of KaiA binding to the $k$th KaiC hexamer, and the term $p_k^B(t) = \sum_{i=1}^{6}\tanh(i/n_B)\sum_{j=0}^{i}P_{C_6B_iA_{2j}}(k,t)$ is the binding level of KaiB. We use $n_B = 1$ in this paper. The term $q_k(t)$ represents the effect of ATPase reactions.

The term $c_1 D_k(t)$ with $c_1 > 0$ in Eq. 7 brings a negative feedback effect in the system. KaiC structure tends to stay in the loose structure with large $D_k$ and small $X_k$, which reduces the binding affinity of KaiA in Eq. 5 and decreases $P_{C_6A_2}(k,t)$ in Eq. 2. As a result, $D_k$ is reduced in Eq. 6. The term $c_3 p_k^B(t)$ with $c_3 > 0$ represents a positive feedback. Large $p_k^B(t)$ value, along with small $X_k$, brings KaiC structure to the loose state, which in turn increases the binding affinity of KaiB to KaiC through Eq. 5 and increases $P_{C_6B_iA_{2j}}(k,t)$ in Eq. 3. Competition among multiple feedback interactions leads the system to have multiple stationary states as described above. However, the system is easily shifted among the states when small perturbations are added to produce oscillation in the system. The term $q_k(t)$, representing the effect of ATPase reactions, works as a perturbation.

We assume that ATPase reactions occur randomly in individual KaiC hexamers; ATP is randomly hydrolyzed with a frequency $f_0$ in each CI domain of subunits of each KaiC hexamer. Biochemical and NMR data suggested that the ATPase activity in the CI is necessary for the binding of KaiC to KaiB [31, 33, 34], which indicated that the ATP hydrolysis enhances the binding affinity of KaiC to KaiB. Therefore, we assume that the ATP hydrolysis changes the KaiC structure from tight to loose states. We represent this effect by writing $q_k(t) = \sum_{i=1}^{6} q(i;k,t)$ with $q(i;k,t) = q_0 > 0$ for $t_0 \le t \le t_0 + \delta_k$ when ATP is hydrolyzed to be ADP and inorganic phosphate (P$_i$) at time $t = t_0$ in the $i$th subunit of the $k$th KaiC hexamer. ADP is kept bound to the subunit for a time duration $\delta_k$ and then released from the subunit. Because the observed ADP release is more frequent in the P-phase with a smaller $\delta_k$ [17], we assume that $\delta_k$ depends on the hexamer structure as

$$\delta_k = \delta_0 (1 - X_k). \qquad (8)$$

For simplicity, we assume that $q(i;k,t) = 0$ when ATP is bound or no nucleotide is bound on the CI domain of the $i$th subunit. Because ATP hydrolysis takes place predominantly in the CI domain and P/dP reactions occur in the CII domain, communication between CI and CII is needed to explain the observed effects of ATP hydrolysis on the P/dP reactions. In the present MM model, allosteric structural change represented by the change in $X_k$ in Eqs. 5-8 enables such communication.

We should note that the following alternative assumption is also possible; $q(i;k,t) = q_0 > 0$ when P$_i$ is released from the KaiC subunit but ADP is kept bound and $q(i;k,t) = 0$ when ATP is bound, ADP and P$_i$ are bound or no nucleotide is bound on the CI [29]. There is no mathematical difference between this assumption and the assumption made in the last paragraph with the present simplified description of the ATPase reactions in the MM model. More precise modeling is necessary for distinguishing these different reaction schemes, but we do not go into these details in the present paper.

Eqs. 2-7 are numerically solved to follow the temporal change of $D_k, X_k, P_{C_6A_2}(k,t)$, $P_{C_6B_iA_{2j}}(k,t)$, and $P_{CBA}(k,t) = \sum_{i=1}^{6}\sum_{j=1}^{i} P_{C_6B_iA_{2j}}(k,t)$ by stochastically varying $q(i;k,t)$ with the frequency $f_0$ and the lifetime $\delta_k$ with $i = 1,\ldots,6$ and $k = 1,\ldots,N$. Oscillations in the ensemble level are monitored by calculating $\bar{D} = \frac{1}{N}\sum_{k=1}^{N} D_k$, $\bar{X} = \frac{1}{N}\sum_{k=1}^{N} X_k$, $\bar{P}_{C_6A_2} = \frac{1}{N}\sum_{k=1}^{N} P_{C_6A_2}(k,t)$, $\bar{P}_{C_6B_iA_{2j}} = \frac{1}{N}\sum_{k=1}^{N} P_{C_6B_iA_{2j}}(k,t)$, and $\bar{P}_{CBA} = \frac{1}{N}\sum_{k=1}^{N} \bar{P}_{CBA}(k,t)$.

## Parameters in the MM model.

We assume that the P/dP reactions are slow enough with the rates $k_p \sim$ 1/(1-2 hour) and $k_{dp} \sim$ 1/(1-2 hour) to realize the period ~24 hour of the P/dP rhythm. The time scale of ATPase reactions, $f_0^{-1}$ and $\delta_k$, should be also as slow as $f_0 \sim$ 1/(0.5-1 hour) and $\delta_k \sim$ 0.5-1 hour; these slow rates of ATPase reactions are consistent with the structural observation [16]. We assume that the binding/dissociation rates of KaiA and KaiB defined as $h_A A^2$, $h_{BA} A^2$, $h_B B$, $f_A$, $f_{BA}$, and $f_B$ should be 1/(minutes) -1/(hours). Such slow binding/dissociation of KaiB is consistent with the observed data [21, 35]. With this rough estimation of parameter values, the ensemble oscillation in the MM model appears to be stable for the wide range of parameter values. An example parameter set is summarized in Table 1. Unless otherwise mentioned, we discuss the results calculated with this parameter set in Results and Discussion section.

**Table 1. Parameters in the MM model.**

| | | | | | |
|---|---|---|---|---|---|
| Rate constant of KaiA binding to KaiC | $h_A^1$ | $6.6 \times 10^{-6}$ h$^{-1}$ | Max. lifetime of ADP binding after hydrolysis | $\delta_0$ | 2.6 h |
| Rate constant of KaiB binding to KaiC | $h_B^0$ | $2 \times 10^{-6}$ h$^{-1}$ | Effect of ATP hydrolysis on structure | $q_0$ | $k_B T_0$ * |
| Rate constant of KaiA binding to KaiB | $h_{BA}$ | $1.8 \times 10^{-6}$ h$^{-1}$ | Base-line temperature effect on structure | $c_0$ | $10 k_B T_0$ * |
| Rate constant of KaiA dissociation from KaiC | $f_A^0$ | $6 \times 10^{-1}$ h$^{-1}$ | Effect of phosphorylation on structure | $c_1$ | $8 k_B T_0$ * |
| Rate constant of KaiB dissociation from KaiC | $f_B^1$ | $4 \times 10^{-1}$ h$^{-1}$ | Effect of KaiA binding on structure | $c_2$ | 0 |
| Rate constant of KaiA dissociation from KaiB | $f_{BA}$ | $2 \times 10^{-1}$ h$^{-1}$ | Effect of KaiB binding on structure | $c_3$ | $4 k_B T_0$ * |
| Constant for variable confinement | $a$ | 2.2 h$^{-1}$ | Total copy number of KaiA monomers | $A_T$ | $2N$ |
| Rate constant of phosphorylation | $k_p$ | $4.4 \times 10^{-1}$ h$^{-1}$ | Total copy number of KaiB monomers | $B_T$ | $20N$ |
| Rate constant of dephosphorylation | $k_{dp}$ | $4.4 \times 10^{-1}$ h$^{-1}$ | Total copy number of KaiC hexamers | $N$ | 1000 |
| Frequency of ATP hydrolysis | $f_0$ | 2.2 h$^{-1}$ | System volume | $V$ | 1 |

*$T_0 = 30$ °C

## The single-molecule (SM) model.

In the MM model, the structure and reactions of individual molecules were described in a simplified manner to facilitate the calculation of an ensemble of a large number of molecules. In order to further analyze the role of ATPase reactions, we introduce a single-molecule model of KaiC hexamer by describing the more details of intra-molecular reactions and structural changes.

In the SM model, each subunit of a hexamer follows two conformational states; $x_i = 0$ when the $i$th subunit is in the loose structural state and $x_i = 1$ when it is in the tight structural state with $i = 1, \ldots, 6$. Though both CI and CII domains have nucleotide binding sites, we focus on the CI domain because ATPase activity is predominant in the CI domain. The four nucleotide binding states in the CI of the $i$th subunit are represented as $a_i = 0$ when no nucleotide is bound, $a_i = 1$ when ATP is bound, $a_i = 2$ when ADP and P$_i$ are bound, and $a_i = 3$ when P$_i$ is released and ADP is kept bound. As shown in Fig. 1b, we define variables, $Ser_i$ and $Thr_i$, as $Ser_i = 1$ ($Thr_i = 1$) when Ser431 (Thr432) of the $i$th subunit is phosphorylated and $Ser_i = 0$ ($Thr_i = 0$) when Ser431 (Thr432) of the $i$th subunit is dephosphorylated. We write $KaiA = 1$ when a KaiA dimer binds to the CII ring and $KaiA = 0$ when a KaiA dimer dissociates from the CII ring. We write $KaiB_i = 1$ when a KaiB monomer binds to the $i$th subunit and $KaiB_i = 0$ when a KaiB monomer dissociates from the $i$th subunit. Thus, as shown in Fig. 1b, the state of KaiC subunits in the SM model is described by the variable $KaiA$ and others;

$$\text{State of KaiC subunits} = KaiA, \{KaiB_i, a_i, Ser_i, Thr_i, x_i\} \text{ with } i = 1, \ldots, 6. \tag{9}$$

Because the atomic structure is rapidly changing with the time scale of milliseconds or shorter but other variables should change more slowly in minutes or longer time scales, we describe the free energy $G$ of the KaiC hexamer as a function of $\{x_i\}$ as $G = G(x_1, x_2, \ldots, x_6)$;

$$G(x_1, x_2, \ldots, x_6) = \sum_{i=1}^{6} x_i \left[ b_0 + b_1^{Thr} Thr_i + b_1^{Ser} Ser_i - b_2 \, KaiA \right.$$
$$\left. + b_3 \, KaiB_i - b_4 \, \theta(a_i) \right] - J \sum_{i=1}^{6} \left( x_i - \frac{1}{2} \right) \left( x_{i+1} - \frac{1}{2} \right), \tag{10}$$

where $b_0, b_1^{Thr}, b_1^{Ser}, b_2, b_3, b_4$, and $J$ are constants. If we regard the fast varying quantities $\{x_i\}$ as "spin variables", $G$ of Eq. 10 should correspond to the "effective Hamiltonian" for spins, whereas the slow variables, $Thr_i, Ser_i, KaiA, KaiB_i$ and $\theta(a_i)$ correspond to the magnetic fields acting on the spins. In other words, phosphorylation states, binding states, and ATPase states give biases for the structural change of each KaiC subunit. $J$ represents coupling between two neighbor subunits; $J > 0$ implies that subunits change their structures in a cooperative way to give rise to two-state allosteric structure transition of the KaiC hexamer. The term $x_i b_4 \theta(a_i)$ represents coupling between ATPase reactions and structure. Here, we assume that $\theta(a_i) = -1$ when $a_i = 2$ or 3, $\theta(a_i) = 1$ when $a_i = 1$, and $\theta(a_i) = 0$ when $a_i = 0$. These assumptions imply that ATP binding increases the tendency of KaiC hexamer structure to be tight while the binding of ADP+Pi or ADP loosens the structure.

For the time scale of minutes, we can represent the structural state of each KaiC subunit with a quasi-equilibrium thermal average of $x_i$ as $\langle x_i \rangle = \text{Tr}\, x_i \exp(-\beta G)/Z$ by approximately treating $Thr_i$, $Ser_i$, $KaiA$, $KaiB_i$ and $\theta(a_i)$ as stationary variables, where Tr represents a sum over all possible values of $\{x_i\}$ and $Z$ is a partition function, $Z = \text{Tr}\exp(-\beta G)$. Then, we represent the hexamer state by using the following variables;

$$Ser = \frac{1}{6}\sum_{i=1}^{6} Ser_i, \quad Thr = \frac{1}{6}\sum_{i=1}^{6} Thr_i,$$

$$X = \frac{1}{6}\sum_{i=1}^{6}\langle x_i \rangle, \quad D = \frac{1}{2}(Ser + Thr), \quad (11)$$

where $X$ and $D$ correspond to $X_k$ and $D_k$ in the MM model.

By comparing Eqs. 7 and 10, we find that $b_0$, $b_1^{Thr}$ and $b_1^{Ser}$, $b_2$, $b_3$, and $b_4$ in the SM model correspond to $c_0$, $c_1$, $c_2$, $c_3$, and $q_0$ in the MM model, respectively. Thus, we find multiple feedback relations working through Eq.10 in the SM model. When $b_2 > 0$, binding of KaiA brings KaiC structure to the tight state as it decreases the energy $G$, which increases the binding affinity of KaiA to KaiC and enhances phosphorylation. When $b_3 > 0$, binding of KaiB brings the structure of KaiC into the loose state. This effect is consistent with the mass spectrometry observation [36] and enhances dephosphorylation, which further increases the binding affinity of KaiB. Therefore, $b_2 > 0$ and $b_3 > 0$ give positive feedback effects in the system. On the other hand, when $b_1^{Ser} > 0$, phosphorylation of Ser431 stabilizes the loose structure of the KaiC subunit by lowering the energy $G$ and this reduces the binding affinity of KaiA and stimulates dephosphorylation of Ser431. Hence, the $b_1^{Ser} > 0$ term provides a negative feedback effect in the system. We assume $b_1^{Thr} < b_1^{Ser}$ so as to make the phosphorylation of Thr432 faster than the phosphorylation of Ser431 as observed in experiment [12, 13]. In the present paper, we assume $b_1^{Thr} < 0 < b_1^{Ser}$. Therefore, the $b_1^{Thr} < 0$ term gives a positive feedback effect. As a result of the coexistence of these multiple feedback relations, the system has multiple metastable states including the phosphorylated loose-structure state and the dephosphorylated tight-structure state. This is why the system tends to stay one of these multiple states and makes transitions among them. ATPase reactions add perturbation to the structure and stimulate the transition between them. The term $x_i b_4 \theta(a_i)$ in Eq. 10 represents such perturbation through the ATPase reactions.

The binding/dissociation reactions of KaiA occur in the CII ring of KaiC hexamer. The reactions are defined as

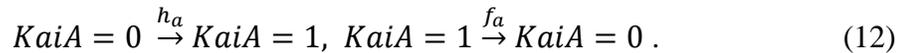

$$KaiA = 0 \xrightarrow{h_a} KaiA = 1, \quad KaiA = 1 \xrightarrow{f_a} KaiA = 0. \quad (12)$$

We consider that the binding affinity of KaiA is regulated by the allosteric structure change of KaiC hexamer; binding of KaiA is enhanced when the hexamer structure is tight whereas binding of KaiA is suppressed when the structure is loose. Therefore, we assume that the rates in Eq. 12 are regulated by $X$ as

$$h_a = h_a^1 X, \qquad f_a = f_a^0(1 - X). \tag{13}$$

A constant $h_a^1$ should depend on the concentration of free unbound KaiA molecules. Here in the present SM model, we use a simplified assumption that the free KaiA concentration is constant, so that $h_a^1$ is kept constant. This assumption represents the situation that the KaiA concentration is much larger than the KaiC concentration, so that the free KaiA concentration is not affected by the oscillation of individual KaiC molecules. In the MM model, as will be shown in Results and Discussion section, sequestration of KaiA to $C_6B_iA_{2j}$ and the resultant temporal depletion of free KaiA concentration is the mechanism of synchronization among multiple KaiC hexamers; synchronization is lost when the free KaiA concentration is kept constant. Because synchronization is not discussed with the SM model, we here adopt a simple assumption that $h_a^1$ is kept constant.

The binding/dissociation reactions of KaiB in the CI domain of KaiC hexamer are described as

$$KaiB_i = 0 \xrightarrow{h_{b_i}} KaiB_i = 1, \quad KaiB_i = 1 \xrightarrow{f_{b_i}} KaiB_i = 0. \tag{14}$$

We assume that the rates in Eq. 14 are regulated by the structure of subunits $\langle x_i \rangle$ in the following manner;

$$h_{b_i} = h_b^0(1 - \langle x_i \rangle), \qquad f_{b_i} = f_b^1 \langle x_i \rangle, \tag{15}$$

In this way, the binding affinity of KaiB to the $i$th subunit of KaiC is decreased when the structure of the subunit is in the tight structure, and the KaiB binding affinity is large when the subunit is in the loose structure. Again, we use a simplified assumption that $h_b^0$ is kept constant through the simulation time course.

The P/dP reactions are described by

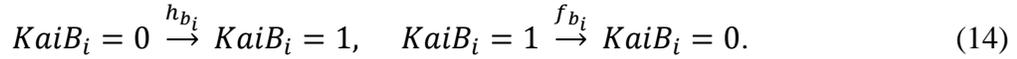

$$Ser_i = 0 \xrightarrow{k_{i,0\to1}^{Ser}} Ser_i = 1, \quad Ser_i = 1 \xrightarrow{k_{i,1\to0}^{Ser}} Ser_i = 0,$$

$$Thr_i = 0 \xrightarrow{k_{i,0\to1}^{Thr}} Thr_i = 1, \quad Thr_i = 1 \xrightarrow{k_{i,1\to0}^{Thr}} Thr_i = 0. \tag{16}$$

The rates in Eq. 16 are regulated by the binding status of KaiA as

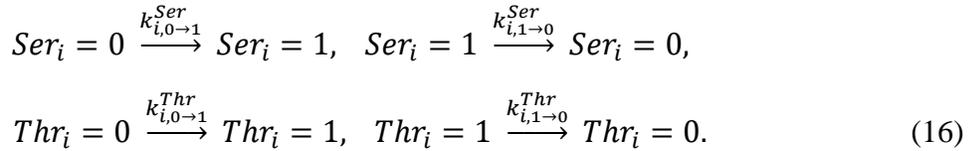

$$k_{i,0\to1}^{Ser} = k_{0\to1}^{Ser,1} KaiA + k_{0\to1}^{Ser,0}(1 - KaiA),$$

$$k_{i,1\to0}^{Ser} = k_{1\to0}^{Ser,1} KaiA + k_{1\to0}^{Ser,0}(1 - KaiA),$$

$$k_{i,0\to1}^{Thr} = k_{0\to1}^{Thr,1} KaiA + k_{0\to1}^{Thr,0}(1 - KaiA),$$

$$k_{i,1\to0}^{Thr} = k_{1\to0}^{Thr,1} KaiA + k_{1\to0}^{Thr,0}(1 - KaiA). \tag{17}$$

We assume the following six equations for the change of nucleotide binding state;

$$a_i = 0 \xrightarrow{k^a_{i,01}} a_i = 1, \quad \text{ATP binding}$$

$$a_i = 1 \xrightarrow{k^a_{i,10}} a_i = 0, \quad \text{ATP unbinding}$$

$$a_i = 1 \xrightarrow{k^a_{i,12}} a_i = 2, \quad \text{hydrolysis reaction}$$

$$a_i = 2 \xrightarrow{k^a_{i,21}} a_i = 1, \quad \text{backward reaction}$$

$$a_i = 2 \xrightarrow{k^a_{i,23}} a_i = 3, \quad P_i \text{ release}$$

$$a_i = 3 \xrightarrow{k^a_{i,30}} a_i = 0. \quad \text{ADP release} \quad (18)$$

Here, both the forward and backward reactions are considered for binding/unbinding of ATP and ATP hydrolysis. Binding reaction of $P_i$ or ADP is not considered by assuming that their concentrations are small. The rates in Eq. 18 are assumed to depend on the structure of each subunit $\langle x_i \rangle$ as in the following manner;

$$k^a_{i,01} = k^{a,1}_{01} \langle x_i \rangle + k^{a,0}_{01} (1 - \langle x_i \rangle),$$

$$k^a_{i,10} = k^{a,1}_{10} \langle x_i \rangle + k^{a,0}_{10} (1 - \langle x_i \rangle),$$

$$k^a_{i,12} = k^{a,1}_{12} \langle x_i \rangle + k^{a,0}_{12} (1 - \langle x_i \rangle),$$

$$k^a_{i,21} = k^{a,1}_{21} \langle x_i \rangle + k^{a,0}_{21} (1 - \langle x_i \rangle),$$

$$k^a_{i,23} = k^{a,1}_{23} \langle x_i \rangle + k^{a,0}_{23} (1 - \langle x_i \rangle),$$

$$k^a_{i,30} = k^{a,1}_{30} \langle x_i \rangle + k^{a,0}_{30} (1 - \langle x_i \rangle). \quad (19)$$

We perform numerical simulations of the stochastic dynamics with SM model: By using $\langle x_i \rangle$ and $X$ calculated with Eq. 10, the rates of reactions are calculated with Eqs. 12-19. $KaiA$, $KaiB_i, a_i, Ser_i,$ and $Thr_i$ are updated with these rates by the Gillespie algorithm [37]. Then, by using thus updated $KaiA$, $KaiB_i, a_i, Ser_i,$ and $Thr_i$, the structural variables $\langle x_i \rangle$ and $X$ in the new time step is calculated with Eq. 10.

**Parameters in the SM model.**

Parameters in the SM model were set to have the same order of magnitude as the corresponding parameters in the MM model. In the SM model, more detailed reaction schemes are considered, so that the larger number of parameters are defined than in the MM model. For example, the P/dP reactions are considered at two sites, Ser431 and Thr432, which are regulated by the KaiA binding status. Therefore, the rates of the P/dP reactions are defined by two sets of parameters, those in the KaiA bound state, $k_{0\to1}^{Ser,1}, k_{1\to0}^{Ser,1}, k_{0\to1}^{Thr,1}, k_{1\to0}^{Thr,1}$, and those in the KaiA unbound state, $k_{0\to1}^{Ser,0}, k_{1\to0}^{Ser,0}, k_{0\to1}^{Thr,0}, k_{1\to0}^{Thr,0}$. Here, superscripts 1 and 0 imply the rates in the KaiA bound and unbound states, respectively. We consider that the phosphorylation rates are larger in the KaiA bound state and the dephosphorylation rates are larger in the KaiA unbound state. Therefore, $k_{0\to1}^{Ser,1}, k_{0\to1}^{Thr,1} > k_{0\to1}^{Ser,0}, k_{0\to1}^{Thr,0}$ and $k_{1\to0}^{Ser,1}, k_{1\to0}^{Thr,1} < k_{1\to0}^{Ser,0}, k_{1\to0}^{Thr,0}$. Because the P/dP reactions at Thr432 precedes the P/dP reactions at Ser431, we consider $k_{0\to1}^{Thr,1} > k_{0\to1}^{Ser,1} > k_{1\to0}^{Thr,1} > k_{1\to0}^{Ser,1}$ and $k_{1\to0}^{Thr,0} > k_{1\to0}^{Ser,0} > k_{0\to1}^{Thr,0} > k_{0\to1}^{Ser,0}$. We assume ATP hydrolysis rates depend on the structure of each subunit; the reaction rates are larger when the structure is tight. Therefore, we assume $k_{01}^{a,1}, k_{10}^{a,1}, k_{12}^{a,1}, k_{21}^{a,1}, k_{23}^{a,1}$, and $k_{30}^{a,1}$ are larger than $k_{01}^{a,0}, k_{10}^{a,0}, k_{12}^{a,0}, k_{21}^{a,0}, k_{23}^{a,0}$, and $k_{30}^{a,0}$. Here, superscripts 1 and 0 indicate the rates in the tight and loose structures, respectively. An example parameter sets, the parameters of rate constants and the parameters in the free energy to determine the structure, are summarized in Tables 2 and 3.

**Table 2. Rate constants in the SM model.***

| | | | | | |
|---|---|---|---|---|---|
| ATP hydrolysis reactions in the tight structure | $k_{01}^{a,1}$ | 1.1 | P/dP reactions with KaiA unbound from KaiC | $k_{0\to1}^{Ser,0}$ | 0.3 |
| | $k_{10}^{a,1}$ | 1.0 | | $k_{1\to0}^{Ser,0}$ | 3.9 |
| | $k_{12}^{a,1}$ | 1.1 | | $k_{0\to1}^{Thr,0}$ | 0.01 |
| | $k_{21}^{a,1}$ | 1.0 | | $k_{1\to0}^{Thr,0}$ | 0.5 |
| | $k_{23}^{a,1}$ | 1.1 | P/dP reactions with KaiA bound to KaiC | $k_{0\to1}^{Ser,1}$ | 4.8 |
| | $k_{30}^{a,1}$ | 1.1 | | $k_{1\to0}^{Ser,1}$ | 0.2 |
| ATP hydrolysis reactions in the loose structure | $k_{01}^{a,0}$ | 0.55 | | $k_{0\to1}^{Thr,1}$ | 0.6 |
| | $k_{10}^{a,0}$ | 0.5 | | $k_{1\to0}^{Thr,1}$ | 0.1 |
| | $k_{12}^{a,0}$ | 0.55 | Binding and dissociation rates of KaiA | $h_a^1$ | 1.5 |
| | $k_{21}^{a,0}$ | 0.5 | | $f_a^0$ | 0.1 |
| | $k_{23}^{a,0}$ | 0.55 | Binding and dissociation rates of KaiB | $h_b^0$ | 1.5 |
| | $k_{30}^{a,0}$ | 0.55 | | $f_b^1$ | 0.2 |

*Values are in units of $h^{-1}$.

**Table 3. Free energy parameters in the SM model.***

| Base-line temperature effect | $b_0$ | 0.13 | Effect of KaiB binding | $b_3$ | 1 |
|---|---|---|---|---|---|
| Effect of Thr432 phosphorylation | $b_1^{Thr}$ | −1 | Effect of ATP hydrolysis | $b_4$ | 4 |
| Effect of Ser431 phosphorylation | $b_1^{Ser}$ | 6 | Coupling between subunits | $J$ | 4.5 |
| Effect of KaiA binding | $b_2$ | 1.5 | | | |

*Values are in units of $k_B T_0$ with $T_0 = 30$ °C.

## Results and Discussion
**Oscillation and synchronization in the ensemble of KaiABC molecules**

In Fig. 2a, we show an example of the ensemble-level oscillation simulated with the MM model. Shown are the KaiC phosphorylation level, $\bar{D}(t)$, structure of KaiC hexamers, $\bar{X}(t)$, probability of $C_6$ of forming the $C_6A_2$ complex, $\bar{P}_{C_6A_2}$, and probability of forming the $C_6B_iA_{2j}$ complexes, $\bar{P}_{CBA}(t)$, where overbars represent that averages were taken over the ensemble of $N$ hexamers. Here, we find that $\bar{P}_{C_6A_2}(t)$ and $\bar{P}_{CBA}(t)$ show counter-phased oscillation. When $\bar{P}_{C_6A_2}(t)$ dominates, binding of KaiA to KaiC hexamer enhances phosphorylation, which increases $\bar{D}$ and decreases $\bar{X}$. On the contrary, when $\bar{P}_{CBA}(t)$ is large, KaiA is sequestered into $C_6B_iA_{2j}$, which depletes the free unbound KaiA. This depletion of free KaiA concentration enhances dephosphorylation, which then decreases $\bar{D}$ and increases $\bar{X}$. This sequestration of KaiA is the mechanism through which individual KaiC hexamers communicate with each other.

Individual oscillation of single KaiC hexamer is shown in Fig. 2b, where oscillation trajectories of $D_k(t)$, $X_k(t)$, $P_{C_6A_2}(k,t)$ and $P_{CBA}(k,t)$ of a KaiC hexamer arbitrarily chosen from the ensemble are plotted. When $X_k(t)$ is large, the lifetime of bound ADP, $\delta_k$, is short and ADP is frequently released from the CI. This short residence of ADP at CI does not affect $X_k(t)$ much. However, when $X_k(t)$ begins to be small, $\delta_k$ becomes longer and ATP hydrolysis provides larger influence to $X_k(t)$. This gives rise to spike-like changes in $X_k(t)$, which induces noisy fluctuating decrease of $D_k(t)$. Synchronization of a large number of KaiC hexamers smooths out this fluctuation to give rise to a coherent oscillation of $\bar{D}$.

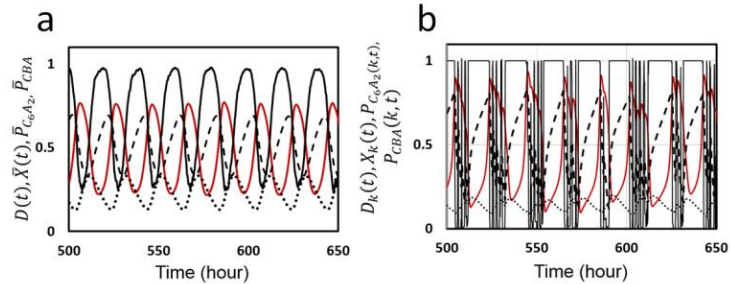

**Figure 2.** Ensemble and single molecule oscillations in the KaiABC system simulated with the MM model. **a**) The ensemble oscillation of the phosphorylation level $\bar{D}(t)$ (red line), structure of KaiC hexamer $\bar{X}(t)$ (black real line), probability to form complex $C_6A_2$, $\bar{P}_{C_6A_2}$ (dashed line), and probability to form complexes $C_6B_iA_{2j}$, $\bar{P}_{CBA}$ (dotted line), are plotted as functions of time. Averaged over the ensemble of $N = 1000$ hexamers. **b**) The single molecule oscillation of phosphorylation level $D_k(t)$ (red line), structure $X_k(t)$ (black real line), $P_{C_6A_2}(k,t)$ (dashed line), $P_{CBA}(k,t)$ (dotted line) of a KaiC hexamer arbitrarily chosen from the ensemble are plotted as functions of time. Parameters in Table 1 were used.

To analyze this synchronization furthermore, we plot individual oscillations $D_k(t)$ of five KaiC hexamers arbitrarily chosen from $N$ hexamers along with the ensemble-averaged oscillation $\overline{D}(t)$. In Fig. 3a, individual and ensemble oscillations calculated with the same parameter values as used in Fig. 2 are shown. Fluctuating individual oscillations are synchronized and entrained into the ensemble oscillation, which gives rise to a stable coherent oscillation of the ensemble. The essential role of KaiA sequestration in this synchronization is clearly shown in Fig. 3b, in which free KaiA concentration is fixed to be $A = 0.2A_T$. When the free KaiA concentration is fixed to be constant as in Fig. 3b, the sequestration mechanism does not work and individual KaiC hexamers lose synchronization among them. Then, the ensemble-level oscillation vanishes though individual molecules remain oscillating with the large amplitude. Thus, the temporal change of the binding affinity of KaiA to KaiC due to the dynamical change of free KaiA concentration is essential for maintaining synchronization. In contrast, as shown in Fig. 3c, synchronization and the ensemble oscillation are maintained when the free KaiB concentration is fixed to be $B = 0.9B_T$. This maintenance of synchronization shows that the temporal change in the concentration of free KaiB is not necessary for synchronization.

Because ATP hydrolysis plays a significant role to generate individual oscillations as shown in Fig. 2b, it is interesting to see how the ATP hydrolysis affects the synchronization. In Fig. 3d, the frequency of ATP hydrolysis, $f_0$, was made smaller from $f_0 = 2.2 \text{ h}^{-1}$ in Fig.3a to $f_0 = 0.8 \text{ h}^{-1}$ in Fig. 3d. We find synchronization is lost with this small frequency of ATP hydrolysis. Therefore, ATP hydrolysis with sufficient frequency is necessary for synchronization. ATP hydrolysis occurring with random timing in individual KaiC subunits should perturb hexamer to adjust to the dynamical oscillation of free KaiA concentration; without this adjustment, individual KaiC hexamers stay oscillating with their own individual phases. As shown in Fig. 3e, by further reducing the ATP hydrolysis frequency to $f_0 = 0.5 \text{ h}^{-1}$, many KaiC hexamers stop oscillating and show small amplitude fluctuation. Therefore, when $f_0$ is decreased, KaiC oscillators first lose synchronization, and then lose individual oscillations.

Together, these results showed that in the MM model KaiA sequestration is a basis for synchronization of individual KaiC hexamers, and ATP hydrolysis helps individual KaiC hexamers to synchronize under the mechanism of KaiA sequestration.

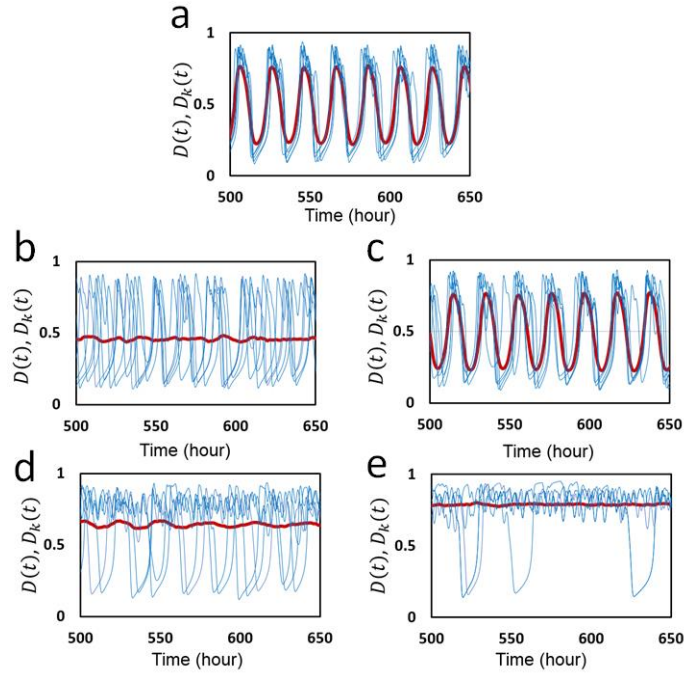

**Figure 3.** Oscillations of individual molecules simulated with the MM model. The phosphorylation level $D_k(t)$ (blue) of five individual KaiC hexamers arbitrarily chosen from the ensemble are superposed with the ensemble average $\bar{D}(t)$ (red). **a)** Parameters in Table 1 were used with $f_0 = 2.2$ h$^{-1}$. **b)** Concentration of free KaiA molecules is fixed to be $A = 0.2A_T$, but concentration of free KaiB molecules remains dynamically changing. **c)** Concentration of free KaiA molecules remains dynamically changing, but concentration of free KaiB molecules is fixed to be $B = 0.9B_T$. Frequency of ATP hydrolysis was made small as **d)** $f_0 = 0.8$ h$^{-1}$ and **e)** $f_0 = 0.5$ h$^{-1}$.

**Temperature compensation.**

An important feature of the KaiABC system is its insensitivity to temperature change, i.e., temperature compensation; the observed $Q_{10}$ defined by the relative change in the oscillation period upon 10 °C temperature change was about 1.06 [1] to $1.01 \pm 0.02$ [38]. However, the physical mechanism of such temperature insensitivity has been elusive. Here, we analyze this problem by using the MM model, particularly by investigating the role of ATP hydrolysis in determining temperature dependence of the oscillation period and amplitude.

The MM model has nine parameters for rate constants, $h_A^1$, $h_B^0$, $h_{BA}$, $f_A^0$, $f_B^1$, $f_{BA}$, $a$, $k_p$, and $k_{dp}$, and two parameters for the ATP hydrolysis reactions, $f_0$ and $\delta_0$. We write their values at temperature $T$ around $T_0 = 30°C$ as

$$h_A^1(T) = [h_A^1(T_0)\exp(\Delta E_{hA}/T_0)] \exp(-\Delta E_{hA}/T),$$

$$h_B^0(T) = [h_B^0(T_0)\exp(\Delta E_{hB}/T_0)] \exp(-\Delta E_{hB}/T),$$

$$h_{BA}(T) = [h_{BA}(T_0)\exp(\Delta E_{hBA}/T_0)] \exp(-\Delta E_{hBA}/T),$$

$$f_A^0(T) = [f_A^0(T_0)\exp(\Delta E_{fA}/T_0)] \exp(-\Delta E_{fA}/T),$$

$$f_B^1(T) = [f_B^1(T_0)\exp(\Delta E_{fB}/T_0)] \exp(-\Delta E_{fB}/T),$$

$$f_{BA}(T) = [f_{BA}(T_0)\exp(\Delta E_{fBA}/T_0)] \exp(-\Delta E_{fBA}/T),$$

$$a(T) = [a(T_0)\exp(\Delta E_a/T_0)] \exp(-\Delta E_a/T),$$

$$k_p(T) = [k_p(T_0)\exp(\Delta E_p/T_0)] \exp(-\Delta E_p/T),$$

$$k_{dp}(T) = [k_{dp}(T_0)\exp(\Delta E_{dp}/T_0)] \exp(-\Delta E_{dp}/T),$$

$$f_0(T) = [f_0(T_0)\exp(\Delta E_{f0}/T_0)] \exp(-\Delta E_{f0}/T),$$

$$\delta_0(T) = [\delta_0(T_0)\exp(-\Delta E_{\delta 0}/T_0)] \exp(\Delta E_{\delta 0}/T). \tag{20}$$

We compare various cases for the values of eleven activation energies, $\Delta E_{hA}$, $\Delta E_{hB}$, $\Delta E_{hBA}$, $\Delta E_{fA}$, $\Delta E_{fB}$, $\Delta E_{fBA}$, $\Delta E_a$, $\Delta E_p$, $\Delta E_{dp}$, $\Delta E_{f0}$, and $\Delta E_{\delta 0}$;

(I) $\Delta E_{hA} = \Delta E_{hB} = \Delta E_{hBA} = \Delta E_{fA} = \Delta E_{fB} = \Delta E_{fBA}$

$$= \Delta E_a = \Delta E_p = \Delta E_{dp} = \Delta E_{f0} = \Delta E_{\delta 0} = 5k_B T_0,$$

(II) $\Delta E_{hBA} = 15k_B T_0$, Other activation energies $= 5k_B T_0$,

(III) $\Delta E_{hBA} = \Delta E_{\delta 0} = 15k_B T_0$, Other activation energies $= 5k_B T_0$,

(IV) $\Delta E_{hBA} = \Delta E_{\delta 0} = 15k_B T_0$, and $\Delta E_{f0} = 0$, Other activation energies $= 5k_B T_0$. (21)

Here, case I is the simplest assumption to make all the relevant activation energies same. Case II is the assumption of the stronger temperature dependence of the binding affinity of KaiA to KaiB. Hatakeyama and Kaneko [20] discussed that enhancement of the KaiA sequestration at higher temperature reduces the phosphorylation rate, which should lead to the temperature compensation. In the present MM model, this enhancement of KaiA sequestration is represented by the larger $\Delta E_{hBA}$ than others. As shown in the previous paper [29], the important features of the MM model are the dependence of the oscillation period on the frequency of the ATP hydrolysis $f_0$ and on the lifetime of the ADP binding state $\delta_0$. In case III, the stronger temperature dependence of $\delta_0$ is assumed with the larger $\Delta E_{\delta 0}$; as temperature increases, the rate of ADP release is enhanced and $\delta_0$ becomes small with the larger temperature dependence than other reactions. The frequency of

ATP hydrolysis was shown to be temperature insensitive [17]. A possible explanation for this temperature insensitivity is the regulation of the reaction rate by the diffusive non-activation type binding process of ATP to the CI, but the precise mechanism of this temperature insensitivity has not been known. Here, to represent this temperature insensitivity, we assume $\Delta E_{f0} = 0$ in case IV.

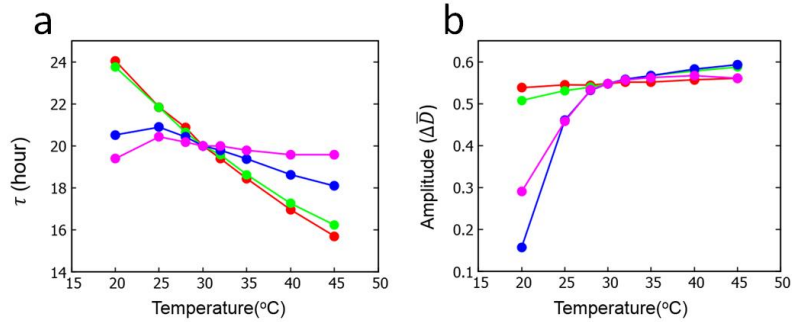

**Figure 4.** Temperature dependence of period and amplitude of the ensemble oscillation $\bar{D}(t)$ simulated with the MM model. **a**) Period τ and **b**) amplitude $\Delta\bar{D}$. Four cases defined in Eq. 21 are compared: case I (red), case II (green), case III (blue), and case IV (magenta). Parameters in Table 1 were used.

In Fig. 4a and Fig. 4b, the calculated temperature dependences of period and amplitude of the ensemble-averaged oscillation $\bar{D}(t)$ are shown for cases I-IV. As shown in Fig. 4a, there is only small difference between case I and case II; therefore, the enhancement of KaiA sequestration at higher temperature contributes only slightly for temperature compensation in the present model. On the other hand, period calculated in cases III and IV is much more insensitive to temperature than period calculated in cases I and II, showing that the enhancement of shortening of the lifetime of the ADP-bound state at higher temperature gives a significant effect on temperature compensation. Temperature insensitivity of ATP hydrolysis frequency gives some effect to make period temperature insensitive as shown in the difference between case III and case IV in Fig. 4a. $Q_{10}$ of the oscillation period calculated between 30 °C and 40 °C is $Q_{10} = 1.17$ (case I), 1.16 (case II), 1.07 (case III), and 1.02 (case IV).

As shown in Fig. 4b, amplitude of oscillation becomes only slightly small as temperature decreases in cases I and II, but amplitude approaches rapidly to 0 as temperature is decreased in cases III and IV. The behavior in cases III and IV is consistent with the observed diminishing of the oscillation amplitude [39], indicating that the oscillation dies out with the Hopf bifurcation mechanism in low temperature.

Summarizing these results, we found that the accelerated release of ADP from the CI with increased temperature largely contributes to temperature compensation of oscillation period, and insensitivity of ATP hydrolysis frequency to temperature gives some help to temperature compensation. The prolonged release of ADP at low temperature is consistent with the observed disappearance [39] of oscillation through the Hopf bifurcation at low temperature. Meanwhile, enhancement of KaiA sequestration does not help much for temperature compensation.

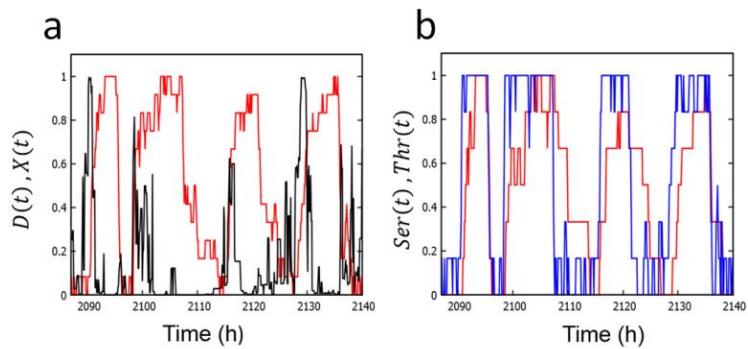

**Figure 5.** Oscillations of single KaiC hexamer simulated with the SM model. **a**) Simulated oscillation of the phosphorylation level $D(t)$ (red) and structure $X(t)$ (black). b) Simulated oscillation of phosphorylation at Ser432 (red) and Thr431 (blue). Parameters in Tables 2 and 3 were used.

**Oscillation in individual KaiC hexamer.**

In the preceding subsections, roles of ATP hydrolysis in the KaiABC oscillation were analyzed with the MM model. Though ATPase reactions randomly occur in individual molecules, they have significant effects on the simulated ensemble-level oscillation. In the following subsections, we use the SM model for further investigating the roles of ATP hydrolysis.

In Fig. 5, we plot the simulated single-molecular phosphorylation levels, $D, Ser$, and $Thr$, and structure $X$; these quantities are defined in Eq. 11 in Method section. As shown in Fig. 5a, the phosphorylation level and structure oscillate with opposite phases in a similar way to the oscillation in the MM model. From the simulated oscillation of $Ser$ and $Thr$ in Fig. 5b, we find that the phase of $Thr$ precedes the phase of $Ser$. In this way, the simulated single-molecular oscillation is noisy stochastic oscillation, but it captures the important features observed in the ensemble-level oscillation. Therefore, the physical origin of features of the ensemble-level oscillation exists at the single-molecular level.

The mechanism of this single-molecular oscillation can be more closely analyzed by changing the parameters, $b_1^{Thr}, b_1^{Ser}, b_2, b_3,$ and $b_4$ in Eq. 10, which represent the strength of coupling between reactions and structural change. In Fig. 6, we plot the oscillation amplitude, $\delta D$, by introducing scaling factors $\alpha_1, \alpha_2,$ and $\alpha_3$ to scale the parameters as $b_1^{Ser} \to \alpha_1 b_1^{Ser}$ (Fig. 6a), $b_1^{Thr}, b_2, b_3 \to \alpha_2 b_1^{Thr}, \alpha_2 b_2, \alpha_2 b_3$ (Fig. 6b), and $b_4 \to \alpha_3 b_4$ (Fig. 6c). Here, $\delta D = 2 \times$(standard deviation of the distribution of $D$), where the values of $D$ were sampled from the trajectory $D(t)$ calculated over $2^{24}$ steps. As discussed in Method section, $\alpha_1$ modulates the strength of negative feedback interaction between P/dP reactions and structure, $\alpha_2$ modulates the strength of positive feedback interactions between P/dP reactions and structure or between binding and structure, and $\alpha_3$ modulates the effect of the ATPase reactions on structure. As shown in Figs. 6a and 6b, the oscillation amplitude decreases when $\alpha_1$ or $\alpha_2$ becomes small, showing that both negative and positive feedback interactions should be sufficiently large for keeping the oscillation amplitude large. As shown in Fig. 6c, the amplitude shows a peak at $\alpha_3 \approx 0.5$. With large $\alpha_3$, the system tends to stay at large $D$ with frequent noisy oscillation of $X$, which prevents regular oscillation, while with small $\alpha_3$, the system tends to stay long at the small $X$ and small $D$ state, which also prevents oscillation. Therefore, there is a suitable range of the coupling strength between ATPase reactions and structure to keep the oscillation amplitude large. To summarize, we find that the large amplitude oscillation in individual KaiC hexamers is supported by negative and positive feedback interactions in structure change as well as the suitable strength of coupling between ATPase reactions and structure change.

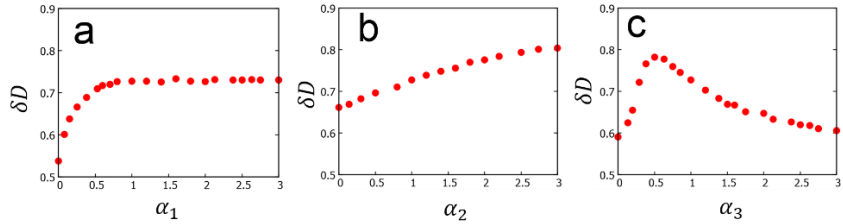

**Figure 6.** Effects of variation of the coupling strength between reactions and structure in the SM model. **a**) Oscillation amplitude $\delta D$ is plotted as a function of the scaling factor $\alpha_1$ of coupling between P/dP reactions at Ser431 and structure. **b**) $\delta D$ as a function of the scaling factor $\alpha_2$ of coupling between P/dP reactions at Thr432 or Kai protein binding reactions and structure. **c**) $\delta D$ as a function of the scaling factor $\alpha_3$ of coupling between ATPase reactions and structure.

**Correlation between ATPase activity and oscillation frequency.**

The observed correlation between the frequency of ATP hydrolysis, or the ATPase activity, and the frequency of KaiC oscillation [16, 17] is important for analyzing the relationship between the ATPase reactions occurring in individual KaiC hexamer and the ensemble-level P/dP rhythm. As discussed in previous subsections, many features of the ensemble-level oscillation are already found in the single-molecular level. Therefore, it is meaningful to analyze the relation between the ATPase activity and the frequency of the single-molecular oscillation.

In Fig. 7a, the single-molecular ATPase activity was calculated with the SM model by introducing a scaling factor $\varepsilon$; the rate constants of ATPase reactions defined in Eq. 19 were uniformly changed from the values of Table 2 as $k_{01}^{a,i} \to \varepsilon k_{01}^{a,i}$, $k_{10}^{a,i} \to \varepsilon k_{10}^{a,i}$, $k_{12}^{a,i} \to \varepsilon k_{12}^{a,i}$, $k_{21}^{a,i} \to \varepsilon k_{21}^{a,i}$, $k_{23}^{a,i} \to \varepsilon k_{23}^{a,i}$, and $k_{30}^{a,i} \to \varepsilon k_{30}^{a,i}$ with $i = 1$ and $0$. With these scaled parameters with various values of $\varepsilon$, the number of released ADP molecules (i.e., the number of $a_i = 3 \to 0$ transitions) was counted as a measure of the ATPase activity in the condition in the absence of KaiA and KaiB, which was represented in the model by imposing $h_a^1 = h_b^0 = 0$. Also shown in Fig. 7a is the averaged peak frequency of the power spectrum of the Fourier transform of the $D$ oscillation calculated with nonzero $h_a^1$ and $h_b^0$ defined in Table 2 for the corresponding values of $\varepsilon$. Because oscillations bear intense stochasticity, the averaged peak frequency and its error bar were derived from extensive calculations: 50 power spectra were calculated from the Fourier transform of 50 trajectories of $D(t)$, with each trajectory having $2^{18}$ data points. Then, the peak of the spectrum obtained by averaging 50 spectra was identified. This calculation was repeated 100 times by using different random numbers, and from the distribution of 100 peak frequencies, the averaged peak frequency was obtained and the error bar was calculated from the standard deviation. In spite of the intense stochasticity as seen from the large error bars in Fig. 7a, we can see that the ATPase activity is correlated to the P/dP oscillation frequency in single KaiC hexamer.

In Fig. 7b, the results of the ensemble calculation with the MM model were plotted. Here, the number of released ADP in the ensemble were calculated in the condition of $A_T = B_T = 0$ with various values of ATP hydrolysis frequency $f_0$, and the corresponding ensemble-level P/dP rhythm was calculated with the nonzero values of $A_T$ and $B_T$ as defined in Table 1. Fig. 7b shows that there is a distinct correlation between ATPase activity and the P/dP rhythm in the ensemble.

Comparing the results of Fig. 7a and Fig. 7b, we find that the important feature of the ensemble-level correlation between the ATPase activity and the P/dP rhythm already appears at the single-molecule level.

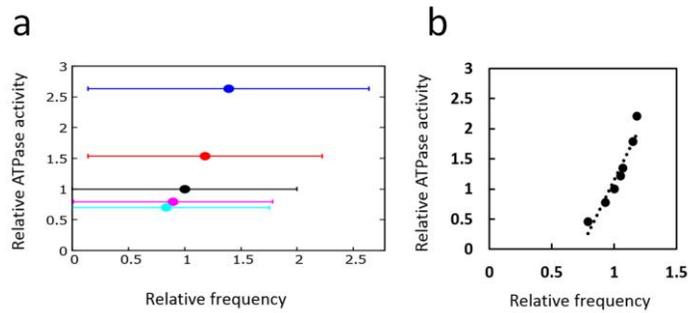

**Figure 7.** Correlation between ATPase activity and frequency of phosphorylation oscillation. **a**) Single-molecular correlation calculated with the SM model. Each point represents the ATPase activity calculated in the non-oscillatory condition and the average frequency calculated in the oscillatory condition for various scaling values $\varepsilon$ of the ATPase reaction rates. Error bars are the standard deviation of the distribution of the peak frequency of the Fourier transform spectra. **b**) Ensemble-level correlation calculated with the MM model. Each point represents the ATPase activity calculated in the non-oscillatory condition and the average frequency calculated in the oscillatory condition for various values of $f_0$. Each point is the average of 10 trajectories, each of which is $2 \times 10^3$ h long.

# Conclusion

We constructed two models, the MM model, which explains oscillation dynamics of an ensemble of a large number of Kai molecules and the dynamics of constituent individual molecules, and the SM model, which explains oscillation dynamics of individual molecules in more details. In these two models, we have hypothesized that KaiC hexamers undergo allosteric transitions between tight and loose structural states. The binding affinity of KaiA and KaiB was assumed to depend on the structure of KaiC hexamer and the rates of P/dP reactions were defined to depend on the binding status of KaiA. The binding of KaiB to KaiC leads to the KaiABC complex formation, which sequesters KaiA to deplete the free KaiA concentration and affects the phosphorylation rate of the other KaiC molecules. ATPase activity of the CI domain of KaiC was assumed to be regulated by the structure of KaiC, and the ATP hydrolysis reactions were assumed to bring about the perturbation on the allosteric transition of the KaiC structure.

Simulations with the MM model showed that stochastic individual oscillations are synchronized through the communication among KaiC hexamers, which was realized by the sequestration of KaiA into the KaiABC complex, and the frequent ATPase reactions in the CI domain perturb individual KaiC hexamers to allow them to generate the ensemble-level oscillation. Thus, both KaiA sequestration and frequent ATPase reactions perform pivotal roles to maintain a coherent ensemble-level oscillation. Temperature dependence of the period and amplitude of the ensemble-level oscillation were examined with the MM model, and it was shown that the intense temperature dependence of the lifetime of the ADP bound state in the CI domain is important to realize the temperature compensation of the oscillation period, and such dependence is consistent with the observed disappearance of oscillation at low temperature [39]. It was also shown that temperature insensitivity of the frequency of ATP hydrolysis helps temperature compensation to some extent. Simulations with the SM model showed that the many features of the ensemble-level oscillation shown with the MM model appear also in the single-molecular oscillation in the SM model though the single-molecular oscillation bears intense stochastic fluctuation. Results in the SM model showed that perturbation of the ATP hydrolysis reactions on the structure is necessary to maintain the large amplitude single-molecular oscillation and that ATPase activity and the P/dP oscillation frequency are correlated to each other at single-molecule level, which largely explains the correlation between ATPase activity and the P/dP oscillation frequency found in the ensemble level.

Thus, both MM and SM models suggest that ATP hydrolysis reactions randomly occurring in individual KaiC hexamers play important roles in synchronization, temperature compensation, and frequency determination at the ensemble-level oscillation. In order to verify these suggestions, the present "mesoscopic" models should be further compared with the microscopic atomic-level observations of structure and reactions [16] and also with the macroscopic ensemble-level observations of phase shift, synchronization, and entrainment [39, 40]. In this way, a comprehensive picture that unifies explanations of microscopic through macroscopic phenomena should open a new perspective of biomolecular systems working coherently in living cell systems.

## Acknowledgements

This work was supported by the Riken Pioneering Project, JST CREST Grant Number JPMJCR15G2, Japan, JSPS KAKENHI Grant Number JP16H02217, and Otsuka Toshimi Scholarship.

## Conflicts of Interest

The authors declare no competing financial interest.

## Author Contributions

S.D., T.P.T., and M.S. designed the project; S.D. and M.S. performed the calculations; S.D., T.P.T., and M.S. analyzed the results; and all authors contributed to writing the manuscript.